\documentclass[journal,twoside]{IEEEtran}

\usepackage{amsmath,amssymb,amsfonts}
\usepackage[final]{graphicx}
\usepackage{psfrag}
\usepackage{epsfig}
\usepackage[numbers,sort&compress]{natbib}
\usepackage{flushend}
\usepackage{array,booktabs}

\newtheorem{lemma}{Lemma}

\makeatletter
\def\blfootnote{\xdef\@thefnmark{}\@footnotetext}
\makeatother

\newcommand*{\bfrac}[2]{\genfrac{}{}{0pt}{}{#1}{#2}}

\def \I {{\mathcal I}}
\def \M {{\mathcal M}}

\begin{document}

\title{{MGF Approach to the Analysis of \\Generalized Two-Ray Fading Models}}
\author{Milind Rao,~\IEEEmembership{Student~Member,~IEEE}, F. Javier Lopez-Martinez,~\IEEEmembership{Member,~IEEE}, Mohamed-Slim Alouini,~\IEEEmembership{Fellow,~IEEE} and Andrea Goldsmith,~\IEEEmembership{Fellow,~IEEE}}


\maketitle
\begin{abstract}
\blfootnote{This work was presented in part at IEEE 48th Annual Conference on Information Sciences and Systems (CISS 2014) \cite{Milind2014}).\\ \indent M. Rao, and A. Goldsmith are with the Wireless Systems Lab, Department of Electrical Engineering, Stanford University, CA, USA. (email: milind@stanford.edu, andrea@wsl.stanford.edu).\\ \indent F. J. Lopez-Martinez was with the Wireless Systems Lab, Department of Electrical Engineering, Stanford University, CA, USA. He is now is with Dpto. Ingenieria de Comunicaciones, Universidad de Malaga, Spain (email: fjlopezm@ic.uma.es). \\ \indent M.-S. Alouini is with Electrical Engineering Program, Computer, Electrical, and Mathematical Science and Engineering (CEMSE) Division, King Abdullah University of Science and Technology (KAUST), Thuwal, Makkah Province, Saudi Arabia. (email: slim.alouini@kaust.edu.sa)\\ \indent
This work was supported by Maitra-Luther fellowship, NEC, Huawei, the NSF Center for Science of Information, the Junta de Andalucia (P11-TIC-7109), Spanish Government-FEDER (TEC2010-18451, TEC2013-44442-P, COFUND2013-40259), the University of Malaga and the European Union under Marie-Curie COFUND U-mobility program (ref. 246550).\\
\indent Copyright (c) 2014 IEEE. Personal use of this material is permitted.  However, permission to use this material for any other purposes must be obtained from the IEEE by sending a request to pubs-permissions@ieee.org.
}
We analyze a class of Generalized Two-Ray (GTR) fading channels that consist of two line of sight (LOS) components with random phase plus a diffuse component. We derive a closed-form expression for the moment generating function (MGF) of the signal-to-noise ratio (SNR) for this model, which greatly simplifies its analysis. This expression arises from the observation that the GTR fading model can be expressed in terms of a conditional underlying Rician distribution. We illustrate the approach to derive simple expressions for statistics and performance metrics of interest such as the amount of fading, the level crossing rate, the symbol error rate, and the ergodic capacity in GTR fading channels. We also show that the effect of considering a more general distribution for the phase difference between the LOS components has an impact on the average SNR.\end{abstract}


\section{Introduction}
We consider a class of fading channels where the fading amplitude is built from two line of sight (LOS) components and multiple non-LOS (NLOS) components. The arriving LOS components can be regarded as individual multipath waves with constant amplitude and random phase, whereas the multiple NLOS components can be grouped into an aggregate diffuse component \cite{durgin2000}. We will denote this general class of fading channels as Generalized Two-Ray (GTR) fading models and specify the phase distribution between the LOS components when used in analysis. 

When uniformly distributed phases for the LOS components are assumed, the resultant GTR fading model (GTR-U) reduces to the Two Wave with Diffuse Power (TWDP) model proposed by Durgin, Rappaport and de Wolf as a generalization of the Rayleigh and Rician fading models \cite{Rap89}. This model was shown to closely match field measurements in indoor scenarios \cite{RapFactory}. By varying the power of the LOS and NLOS components, the TWDP fading model encompasses the Rayleigh and Rician models along with the LOS case with no diffuse components (i.e., a two-ray model). Another fading behavior that TWDP fading can model is when the fading is more severe than Rayleigh fading \cite{hyperR1}. This regime, termed \emph{hyper-Rayleigh} fading, has been observed in wireless sensor networks deployed in cavity structures such as an aircraft or a bus \cite{WhyHyperRal}, or in vehicle-to-vehicle communication links \cite{WorseThanRay}. Other distributions, such as the $\kappa$-$\mu$ extreme distribution, have been proposed to model \emph{hyper-Rayleigh} fading behavior \cite{Extreme}.

Although this fading model can indeed suit a variety of propagation conditions, its complicated statistical characterization has been its main drawback. The original pdf in \cite{Rap89} is given in integral form, which has hindered the wireless system performance analysis using this model. To circumvent this issue, an \textit{approximate} closed-form pdf was also proposed in \cite{Rap89} to facilitate obtaining analytical results for this channel. This approximate TWDP fading pdf has been widely used to characterize the performance of wireless communication systems in TWDP fading, in terms of the bit error rate (BER) in single-antenna and multi-antenna reception using various modulation schemes \cite{BPSKTWDP,MRCTWDP,SCTWDP,QAMTWDP}, as well as in relay networks \cite{RelayTWDP1,RelayTWDP2}. Other performance metrics such as the secrecy capacity associated with physical layer security have also been investigated \cite{TWDPsecrecy}. 

These works have provided the first analytical results for TWDP fading in a number of scenarios. They are, however, approximations, and their accuracy is known to degrade when the two LOS components are very strong and their magnitudes are similar \cite{Rap89}. In particular, the exact characterization of most performance metrics in TWDP fading remains an open problem. This issue was recently addressed in \cite{TWDPalternate}, where alternative \textit{exact} expressions for the TWDP fading pdf and cdf were given in terms of an infinite series of Laguerre and Legendre polynomials.

Interestingly, the authors in \cite{Rap89} posited that the TWDP fading pdf \textit{somewhat resembles} the Rician pdf, but did not further exploit this similarity. We have found that characterizing the envelope statistics of this fading model is closely related to a classical problem in communication theory addressed by Rice \cite{RicSineNoise} on the statistical properties of sine waves in Gaussian noise. Esposito and Wilson \cite{Sin2gauss} further developed these ideas and provided expressions for the distribution of two sine waves in the presence of Gaussian noise. 

Motivated by these results, we show that the envelope statistics of the GTR-U fading model conditioned on the phase difference between the LOS components results in the Rician fading model. This allows us to express \emph{any performance} \emph{metric} that is a linear function of the envelope statistics of the GTR-U fading model in terms of a finite integral over the performance metric for the Rician case. 

As a key result, we obtain a closed-form expression for the Moment Generating Function (MGF), which to the best of our knowledge has not been expressed in the literature so far. With the MGF in closed-form, we can easily analyze the symbol error probability of multi-channel reception schemes \cite{AlouiniBook} as well as evaluate the ergodic capacity \cite{DiRenzo,yilmaz2012} in GTR-U fading. Using this simple yet powerful approach, we also find simple expressions for many statistics of interest such as the pdf, cdf, the amount of fading (AOF) and the level crossing rate (LCR). 

Inspired by the connection between the Rician and GTR-U fading unveiled above, we also show that the statistical properties of the phase difference between the two LOS components $\alpha$ have an impact on the fading experienced by the signal. Allowing this phase difference $\alpha$ to be arbitrarily distributed, we analyze a more general fading propagation condition: the GTR fading model with arbitrary phase. We will show that this additional degree of freedom models a much larger range of fading behavior, and hence can be useful to characterize \textit{hyper-Rayleigh} fading in more severe scenarios than the ones considered in \cite{hyperR1,WhyHyperRal,WorseThanRay}. Interestingly, we also obtain a closed-form expression for the MGF of the GTR fading model when the phase difference is distributed according to the von Mises (or circular normal) distribution \cite{vonMises,Gumbel}, which includes the uniform distribution as a particular case. Hence, the analysis in this new general scenario is of similar complexity to the conventional GTR-U fading case.

The remainder of this paper is structured as follows: in Section \ref{Conn}, we present the connection between the Rician and GTR-U fading models as a key aspect in our analysis. Then, in Section \ref{StatAnalysis} we derive a closed-form expression for the MGF, as well as new expressions for other statistics of the GTR-U model such as the moments, LCR and AOF. In Section \ref{GTR} we discuss the effect of the distribution of the phase difference between the LOS components, introducing the GTR fading model with arbitrary phase as a natural extension of the conventional GTR-U model. Section \ref{MGFSysPer} employs the MGF approach to analyze some system performance metrics in GTR-U fading: the SEP and the ergodic capacity. The implications for system design enabled by our analysis are presented in section \ref{SysTradeOff}. The main conclusions are outlined in Section \ref{conclusion}.

\section{Connections between GTR-U and Rician Fading Models}
\label{Conn}
\subsection{A brief description of the GTR-U fading model}
As presented in \cite{Goldtext}, the complex baseband received signal $s(t)$ in narrowband multipath fading is:
\begin{equation}
s(t)=\Re\left\{ u(t)\underset{n}{\sum}\alpha_{n}e^{j\phi_{n}}\right\},\label{eq:mulsum}
\end{equation}
 where $u(t)$ is the transmitted signal in baseband, $\alpha_{n}$ and $\phi_{n}$ represent the amplitude and phase of the $n$-th multipath
component and $\Re\{.\}$ denotes the real part.

The GTR-U fading model described in \cite[eq. 7]{Rap89} consists of two specular components and a diffuse component, as
\begin{equation}
V_{r}=V_{1}\exp(j\phi_{1})+V_{2}\exp(j\phi_{2})+X+jY,\label{eq:TWDPdefn}
\end{equation}
where $V_{r}$ is the received signal, components $1$ and $2$ are specular components with $\phi_{1},\,\phi_{2}\sim\mathcal{U}(0,2\pi)$ and $V_{1}$ and $V_{2}$ are constant. In the diffuse component $X,\, Y\sim\mathcal{N}(0,\sigma^{2})$. Throughout the rest of the paper, we will refer in the text to the GTR-U model as simply GTR for the sake of brevity. A distinction will be made where necessary to include the effect of a different phase distribution (e.g. in section \ref{GTR}).

The model is conveniently expressed in terms of the parameters $K$ and $\Delta$, defined
as
\begin{align}
\label{eq:Kequiv}
K & = \frac{V_{1}^{2}+V_{2}^{2}}{2\sigma^{2}},\\
\Delta & = \frac{2V_{1}V_{2}}{V_{1}^{2}+V_{2}^{2}}.
\end{align}
Similar to the Rician parameter, here $K$ represents the ratio of the power of the specular components to the diffuse power; $\Delta$ is related to the ratio of the peak specular power to the average specular power and serves as the comparison of the power levels of the two specular components. We observe that $\Delta=1$ only when the two specular components are of equal amplitude, and $\Delta=0$ when either LOS component has zero power. Special cases of the GTR fading model are detailed in \cite{Rap89}, encompassing the \emph{One Wave}, \emph{Two Wave}, Rayleigh and Rician fading models. In \cite{hyperR1} it is shown that when $K>0$ and $\Delta\approx1$ the channel exhibits worse fading than Rayleigh, referred to as hyper-Rayleigh behavior. As $K$ increases, the fading becomes more severe and with the extreme condition of $K\rightarrow\infty$, the most severe two-wave fading model emerges

The pdf of the GTR fading model was given in \cite{Rap89} as 
\begin{align}
f_{\text{GTR-U}}(r)=r\int_{0}^{\infty}e^{-\frac{v^{2}\sigma^{2}}{2}}J_{0}(V_{1}v)J_{0}(V_{2}v)J_{0}(vr)vdv.\label{eq:TWDPlongform}
\end{align}
where $J_0(\cdot)$ denotes the Bessel function of the first kind with order zero. An alternative expression for this pdf was also given as
\begin{align}
\label{eq:TWDPlongform2}
&f_{\text{GTR-U}}(r)=\frac{r}{\sigma^2}\exp{\left(-\frac{r^{2}}{2\sigma^{2}}-K\right)}\times\\&\frac{1}{\pi}\int_{0}^{\pi}\exp{\left(K\Delta\cos\theta\right)}I_{0}\left(\frac{r}{\sigma}\sqrt{2K(1-\Delta\cos\theta)}\right)d\theta,\nonumber
\end{align}
where $I_{0}(\cdot)$ is the modified Bessel function of the first kind with order zero. Since both (\ref{eq:TWDPlongform}) and (\ref{eq:TWDPlongform2}) are in integral form, the authors in \cite{Rap89} presented an approximate representation of the pdf as 
\begin{equation}
f_{\text{GTR-U}}(r)\approx\frac{r}{\sigma^{2}}\exp\left(-\frac{r^{2}}{2\sigma^{2}}-K\right)\sum_{i=1}^{M}a_{i}D\left(\frac{r}{\sigma^{2}};K,\,\Delta\alpha_{i}\right),
\label{eq:TWDPpdfApp}
\end{equation}
where $a_{i}$ are tabulated constants, the order $M$ should be sufficiently large and
\begin{align}
D(x;K,\,\alpha_i)=&\frac{1}{2}\exp(\alpha_i K)I_{0}\left(x\sqrt{2K(1-\alpha_i)}\right)\\&+\frac{1}{2}\exp(-\alpha_i K)I_{0}\left(x\sqrt{2K(1+\alpha_i)}\right)\nonumber,
\end{align}
where $\alpha_i =\cos\left(\frac{\pi(i-1)}{2M-1}\right)$.
\subsection{GTR-U fading as a generalization of Rician fading}

Similar to the procedure followed in \cite{Rap89} to derive (\ref{eq:TWDPlongform2}) from (\ref{eq:TWDPlongform}), we use an expanded form of the Bessel function $J_0$ which results in
\begin{multline}
f_{\text{GTR-U}}(r)=r\int_{0}^{\infty}v\exp(\frac{-v^{2}\sigma^{2}}{2})J_{0}(vr)\frac{1}{(2\pi)^{2}}\\
\times\int_{\theta=0}^{2\pi}\int_{\phi=0}^{2\pi}\exp[jV_{1}v\cos(\theta)+jV_{2}v\cos(\phi)]d\theta d\phi dv\label{eq:longexpTWDP},
\end{multline}
We recognize that
\begin{align}
V_{1}&\cos(\theta)+V_{2}\cos(\phi)=V_{1}\cos(\theta)+V_{2}\cos(\theta-\alpha)\nonumber\\
=&[V_{1}+V_{2}\cos(\alpha)]\cos(\theta)+V_{2}\sin(\alpha)\sin(\theta)\nonumber\\
=&\sqrt{V_{1}^{2}+V_{2}^{2}+2V_{1}V_{2}\cos(\alpha)}\cos(\theta+\theta_{0}),\label{eq:bessSimpl}
\end{align}
where $\alpha=\theta-\phi$ is the phase difference between the two LOS components and $\theta_{0}=\arctan(\frac{V_{2}\sin(\alpha)}{[V_{1}+V_{2}\cos(\alpha)]})$.
Using (\ref{eq:bessSimpl}) in (\ref{eq:longexpTWDP}) and noticing that adding a phase term $\theta_{0}$ in the Bessel function integrand does not affect as it is integrated over an entire period, we get
\begin{multline}
f_{\text{GTR-U}}(r)=\frac{1}{2\pi}\int_{\alpha=0}^{2\pi}r\int_{0}^{\infty}v\exp(\frac{-v^{2}\sigma^{2}}{2})J_{0}(vr)\\
\times J_{0}\left(\sqrt{V_{1}^{2}+V_{2}^{2}+2V_{1}V_{2}\cos(\alpha)}\right)dvd\alpha.\label{eq:stepinRiceTWDPder}
\end{multline}
The inner integral of (\ref{eq:stepinRiceTWDPder}) is seen to be a special case of (\ref{eq:TWDPlongform}) with only one LOS component $\bar{V}_{1}$
and $\bar{V}_{2}=0$, i.e. it can be seen as an \textit{equivalent} Rician pdf. The equivalent LOS component amplitude $\bar{V}_{1}$ is given by, 
\begin{align}
\bar{V}_1 & =\sqrt{V_{1}^{2}+V_{2}^{2}+2V_{1}V_{2}\cos(\alpha)}\label{eq:VRiceEq}\\
\bar{K} &= K\left(1+\Delta\cos(\alpha)\right).\label{eq:kRiceEq}
\end{align}
Employing the equivalent Rician pdf in (\ref{eq:stepinRiceTWDPder}), we obtain
\begin{equation}
f_{\text{GTR-U}}(r)=\frac{1}{2\pi}\int_{0}^{2\pi}f_{\text{Rice}}\left(r;\, K[1+\Delta\cos(\alpha)]\right)d\alpha.\label{eq:TWDPasRice}
\end{equation}

Thus, we see that the pdf of the GTR fading model is obtained by finding the Rician pdf with equivalent $\bar{K}$ as given by (\ref{eq:Kequiv}) and then averaging over $\alpha$, the phase difference between the LOS components. If we plug the well-known expression for the Rician pdf \cite{AlouiniBook} given by
\begin{equation}
\label{eq:Ricepdf}
f_{\text{Rice}}(r)=\frac{r}{\sigma^2}e^{-\frac{r^{2}}{2\sigma^{2}}-K}I_{0}\left(\frac{r}{\sigma}\sqrt{2K}\right)
\end{equation}
in (\ref{eq:TWDPasRice}), we obtain the following expression for the GTR fading pdf
\begin{align}
\label{eq:perfect}
f_{\text{GTR-U}}(r)&=\frac{r}{\sigma^2}e^{-\frac{r^{2}}{2\sigma^{2}}-K}\times\\&\frac{1}{2\pi}\int_{0}^{2\pi}e^{K\Delta\cos(\alpha)}I_{0}\left(\frac{r}{\sigma}\sqrt{2K[1+\Delta\cos(\alpha)]}\right)d\alpha,\nonumber
\end{align}
which is very similar to (\ref{eq:TWDPlongform2}). It is straightforward to show that both are coincident by a simple change of variables.

We have been able to find an insightful connection between the GTR and the Rician fading models, showing that the former is in fact a natural generalization of Rician fading for two LOS components. This connection can be inferred for an arbitrary number of LOS components; however, as discussed in \cite{Rap89}, the applicability of such an $n-$wave model is questionable in practice.

Another intuitive approach to arriving at (\ref{eq:perfect}) is as follows: conditioning the received signal amplitude on the phase difference
between the LOS components we get
\begin{equation}
V_{r}=\exp(j\phi_{1})\left(V_{1}+V_{2}\exp[j(\alpha)]\right)+V_{\text{diff}}.\label{eq:condTWDP}
\end{equation}
This problem is equivalent to finding the Rician pdf as there is a single LOS component of uniformly distributed phase $\phi_{1}$ and constant amplitude $\bar{V}_1$ and $\bar{K}$ given in (\ref{eq:VRiceEq}) and (\ref{eq:kRiceEq}), respectively. In fact, thanks to the circular symmetry of $V_{\text{diff}}$ the envelope statistics of (\ref{eq:condTWDP}) are independent of the distribution of $\phi_1$ and only depend on the phase difference $\alpha$ \cite{Simon85}. Thus, the GTR fading model conditioned on the phase difference $\alpha$ results in the Rician envelope distribution, i.e. 
\begin{equation}
\label{eq:condPDF}
{f_{\text{GTR-U}}(r|\alpha)=f_{\text{Rice}}\big(r;\ K[1+\Delta\cos(\alpha)]\big)}.
\end{equation}
Given that $\phi_{1},\,\phi_{2}\sim\mathcal{U}(0,\,2\pi),$ the random variable ${\alpha=\phi_{2}-\phi_{1}\sim\mathcal{U}(0,\,2\pi)}.$ Although
$\phi_{2}-\phi_{1}$ is a symmetric triangular distribution from $-2\pi$ to $2\pi$, we are interested in the phase difference modulo $2\pi$
and $\alpha$ results in a uniformly distributed pdf. Employing the uniform distribution in (\ref{eq:TWDPconditionalFullRice}), we obtain
(\ref{eq:TWDPasRice}). 

The phase difference $\alpha$ could also arise from any arbitrary distribution with pdf $f_\alpha(.)$ and the preceding analysis holds. This is further described in section \ref{GTR}. The pdf of this GTR fading model with arbitrary phase is given as,
  
\begin{align}
f_{\text{GTR}}(r)&=\int_{0}^{2\pi}f_{\text{Rice}}\left(r;\, K[1+\Delta\cos(\alpha)]\right)f_{\alpha}(\alpha)d\alpha
\end{align}

\section{Statistical Analysis of the GTR-U Model}
\label{StatAnalysis}
In this section we describe how the connection unveiled above between Rician and GTR fading models allows us to calculate performance metrics for the latter. As we will later see, this will enable the derivation of a closed form expression for the MGF of the GTR fading model. The following lemma will be of use: 
\begin{lemma}
\label{lemma1}
\textit{Let $H_{R}(\theta)$ be a general metric of a fading model with parameter $\theta$, expressed as a linear function of its envelope pdf in the form}
\begin{equation}
H_{R}(\theta)=\int_a^b f_{R}(r)g(r)dr,\label{eq:metricLabelTWDPstepinProof}
\end{equation}
\textit{where $0\leq a\leq\ b \leq \infty$ and $g(\cdot)$ is an arbitrary function defined on $\mathbb{R}$. Then, any general metric $H_{\text{GTR}}(K,\Delta)$ of the GTR fading model with parameters $K$, $\Delta$ and phase difference $\alpha$ between the LOS components arising from distribution $f_\alpha(.)$ can be expressed in terms of the same metric of the Rician fading model $H_{\text{Rice}}(K)$ as}
\begin{equation}
H_{\text{GTR}}(K,\Delta)=\frac{1}{2\pi}\int_{0}^{2\pi}H_{\text{Rice}}\big(K[1+\Delta\cos(\alpha)]\big)f_\alpha(\alpha)d\alpha\label{eq:metricTWDP}.
\end{equation}
\end{lemma}

\begin{IEEEproof}
This is easily verified by changing the order of integration in (\ref{eq:metricLabelTWDPstepinProof}). 
\end{IEEEproof}

This simple approach to derive performance metrics and statistics for the GTR fading model is new in the literature to the best of our knowledge. We note that a similar connection has been recently established between Rayleigh and Hoyt (Nakagami-$q$) NLOS fading models in \cite{Romero2014}; however, in the present work the parameter $\alpha$ has a clear and intuitive interpretation as it is related to the phase difference between the two LOS components.

We now apply this lemma to find expressions for some performance metrics of the GTR fading model. 
\subsection{MGF of the GTR-U Fading Model}
\label{MGFSubSec}

The moment generating function (MGF) of the SNR for the Rician fading model is given by 
\begin{equation}
\mathcal{M}_{\text{Rice}}(s)=\frac{1+K}{1+K-s\bar{\gamma}}\exp\left(\frac{Ks\bar{\gamma}}{1+K-s\bar{\gamma}}\right).
\label{eq:mgfRice}
\end{equation}

Lemma \ref{lemma1} holds when the metric is a linear function of the envelope statistics of the fading model. However, most performance metrics (e.g. error probability or capacity) are calculated using the statistics of the SNR instead of the fading amplitude. The average SNR at the receiver is defined as ${\bar{\gamma}=\bar{P}_{r}/N_{0},}$ where $\bar{P}_{r}=V_{1}^{2}+V_{2}^{2}+2\sigma^{2}$ is the average received power and $N_{0}/2$ is the Power Spectral Density of the AWGN noise. Since the average SNR is expressed as
\begin{equation}
\bar{\gamma}=(1+K){2\sigma^{2}}/{N_{0}},
\label{eq:gamma}
\end{equation} 
the pdf of $\gamma$ is given by 
\begin{equation}
f_{\gamma}(\gamma)=\frac{f_{R}\left(\sqrt{\bar{P}_{r}\gamma/\bar{\gamma}}\right)}{2\sqrt{\bar{\gamma}\gamma/\bar{P}_{r}}}.
\label{eq:pdfGamma}
\end{equation}

We see that $\frac{1+K}{\bar{\gamma}}$ is constant both for Rician and GTR fading models and it equals $\frac{N_0}{2\sigma^2}$; hence, it represents the ratio of noise introduced by the receiver to the power of the diffuse component according to (\ref{eq:gamma}). When using a certain performance metric derived for Rician fading to obtain the equivalent metric for GTR fading and the metric of interest is a function of $\bar\gamma$, then $\bar{K}(\alpha)$ should not be substituted in place of $K$ where a term $\frac{1+K}{\bar{\gamma}}$ appears in the equivalent expression for the Rician metric before integration. With these considerations, the MGF of the GTR fading model is calculated using Lemma \ref{lemma1} and the MGF of the Rician model as
\begin{align}
&\mathcal{M}_{\text{GTR-U}}(s)=\frac{1}{2\pi}\int_{0}^{2\pi}\frac{1+K}{1+K-s\bar{\gamma}}\exp\left(\frac{\bar{K}(\alpha)s\bar{\gamma}}{1+K-s\bar{\gamma}}\right)d\alpha\nonumber\\
&=\frac{1+K}{1+K-s\bar{\gamma}}\exp\left(\frac{Ks\bar{\gamma}}{1+K-s\bar{\gamma}}\right)I_{0}\left(\frac{Ks\bar{\gamma}\Delta}{1+K-s\bar{\gamma}}\right).\label{eq:MTWDP}
\end{align}

Hence, we have found a closed-form expression for the MGF of the GTR fading model. Even though the GTR fading pdf cannot be expressed in closed-form, we have shown that the MGF is characterized by a very simple expression. This has two direct implications: first, the moments for the GTR fading model can also be expressed in closed-form, using Leibniz's rule for the derivative of products. Secondly, the MGF is extensively used to characterize performance of digital communication systems \cite{AlouiniBook}. Therefore, expression (\ref{eq:MTWDP}) is useful to analyze some of the scenarios considered in the literature \cite{BPSKTWDP,MRCTWDP,SCTWDP,QAMTWDP} without the need for using the approximate pdf in (\ref{eq:TWDPpdfApp}).

\subsection{Statistics of the GTR-U fading model}
We now use Lemma \ref{lemma1} to obtain simple expressions for other statistics of interest of the GTR fading model. 

\subsubsection{Probability density function}

Using the pdf of the Rician distribution given in (\ref{eq:Ricepdf}), the pdf of the fading envelope for the GTR fading model was given in (\ref{eq:perfect}). From (\ref{eq:pdfGamma}), we find the pdf of the SNR in GTR fading to be 
\begin{align}
&f_{\text{GTR-U}}(\gamma)=\frac{1+K}{\bar{\gamma}}\exp\left\{-\frac{\gamma(1+K)}{\bar{\gamma}}\right\}\frac{1}{2\pi}\nonumber\\
&\times \int_{0}^{2\pi}\exp\left\{-\bar{K}(\alpha)\right\}I_{0}\left(2\sqrt{\frac{\gamma}{\bar{\gamma}}\bar{K}(\alpha)[K+1]}\right)d\alpha,
\end{align}
where $\bar{K}(\alpha)$ is defined in (\ref{eq:kRiceEq}). We observe that the case where $\Delta=0$ reduces to the scenario where $\bar{K}(\alpha)=K$ and the resulting pdf is equivalent to the Rician pdf as expected. Furthermore, taking $K=0,$ we get the exponential distribution that characterizes the SNR distribution of Rayleigh fading. 

\subsubsection{Cumulative distribution function}
The cdf of the Rice distribution is
\begin{equation}
\label{eq:cdfRice}
F_{\text{Rice}}(r)=1-Q_{1}(\sqrt{2K},\frac{r}{\sigma}),
\end{equation}
where $Q_{1}(\cdot,\cdot)$ is the Marcum $Q-$function. Hence, the cdf of the GTR fading model is directly given by
\begin{equation}
\label{eq:cdfTWDP}
F_{\text{GTR-U}}(r)=1-\tfrac{1}{2\pi}\int_{0}^{2\pi}Q_{1}\big(\sqrt{2K[1+\Delta\cos(\alpha)]},\frac{r}{\sigma}\big)\, d\alpha.
\end{equation}

\subsubsection{Moments}
The moments of the SNR in GTR fading model can be directly obtained from the MGF. However, it is also possible to calculate these moments from the moments of the SNR of the Rician distribution, given by
\begin{equation}
\mathbb{E}_{\text{Rice}}(\gamma^{k})=\frac{k!}{(1+K)^{k}}{_{1}}F_{1}(-k,\,1;\,-K)\bar{\gamma}^{k},
\label{eq:}
\end{equation}
where ${_{1}}F_{1}(\cdot,\cdot;\cdot)$ is the Kummer confluent hypergeometric function. Using (\ref{eq:metricTWDP}), we have
\begin{equation}
\mathbb{E}_{\text{GTR-U}}(\gamma^{k})=\frac{k!\bar{\gamma}^{k}}{(1+K)^{k}2\pi}\int_{0}^{2\pi}{_{1}}F_{1}\big(-k,\,1;\,-\bar{K}(\alpha)\big)d\alpha.\label{eq:momentsTWDP}
\end{equation}
An alternative expression in terms of the Laguerre polynomials $L_k(\cdot)$ can also be derived, using the well-known relationship $L_k(z)={}_1F_1(-k;1;z)$.

Specifically, the first two moments are given by
\begin{align}
\mathbb{E}_{\text{GTR-U}}(\gamma)=&~\bar{\gamma},\\
\mathbb{E}_{\text{GTR-U}}(\gamma^{2})=&\frac{\bar{\gamma}^{2}}{(1+K)^{2}}\left\{2+4K+K^{2}\left(1+\frac{\Delta^{2}}{2}\right)\right\}.\label{eq:g2TWDP}
\end{align}
\subsubsection{Amount of Fading}

The Amount of Fading \cite{AlouiniBook} is a simple performance criterion to assess the fading model. It is very useful in the analysis of diversity systems, since it allows us to evaluate the severity of fading by using higher moments of the SNR. This metric is defined as follows
\begin{equation}
AF=\frac{\mathbb{E}[(\gamma-\bar{\gamma})^{2}]}{\mathbb{E}[\gamma]^{2}}.
\label{eq:AFRDef}
\end{equation}
The Amount of Fading in the GTR fading model for which a closed-form expression has hitherto not been found is thus easily seen to be
\begin{equation}
AF_{\text{GTR-U}}=\frac{2+4K+K^{2}\Delta^{2}}{2(1+K)^{2}}.
\label{eq:AFTWDP}
\end{equation}

\subsubsection{Level Crossing Rate}
The Level Crossing Rate (LCR) is extensively used in communication theory as a metric that characterizes the rate of change of a random process. The LCR provides information about how often the fading envelope crosses a specific threshold value $r_{th}$, and admits a general representation in integral form given by 
Rice \cite{RicSineNoise} as
\begin{equation}
N(r_{th})=\int_{0}^{\infty}\dot{r}f_{r,\dot{r}}(r_{th},\dot{r})d\dot{r},
\label{eq:LCRRice}
\end{equation}
where $\dot{r}$ is the time derivative of the fading envelope. In the case where the specular components arrive perpendicular to the direction of motion (they do not undergo Doppler fading) and the diffuse component consists of isotropic $2$-$D$ scattering, it is seen that the fading envelope and its time derivative are independent, i.e.
${f_{r,\dot{r}}(r_{th},\dot{r})=f_{r}(r_{th})f_{r}(\dot{r})}$.

In this scenario, the LCR for the Rician fading envelope is known to be
\begin{equation}
N_{Rice}(r_{th})=\sqrt{\frac{\pi}{2}}\times\sqrt{\frac{\bar{P}_r}{K+1}}f_{D}f_{\text{Rice}}(r_{th}),
\label{eq:LCRRician}
\end{equation}
where $f_{D}$ is the maximum Doppler frequency. Hence, assuming the two LOS components do not experience Doppler fading, the LCR for the GTR fading channel is directly given by
\begin{equation}
N_{\text{GTR-U}}(r_{th})=\sqrt{\frac{\pi}{2}}\times\sqrt{\frac{\bar{P}_r}{K+1}}f_{D}f_{\text{GTR-U}}(r_{th})\label{eq:LCRTWDP}.
\end{equation}
The Average Outage Duration (AOD) is a metric that indicates how long the channel is in a fade level below a certain threshold, and is defined as the quotient between the cdf and the LCR, i.e. $A(r_{th})={\Pr(r<r_{th})}/{N(r_{th})}$. Hence, the AOD for GTR fading is given by
\begin{equation}
A_{\text{GTR-U}}(r_{th}) = \sqrt{\frac{2(K+1)}{\pi \bar{P}_r}}\frac{F_{\text{GTR-U}}(r_{th})}{f_{D}~f_{\text{GTR-U}}(r_{th})}\label{eq:AODTWDP}.
\end{equation}

\section{GTR Fading Models with Arbitrary Phase}
\label{GTR}
\subsection{Physical justification.}

In the previous analysis, the phase difference $\alpha$ between the two LOS components in GTR fading is modeled to be uniformly and independently distributed. With this consideration, the GTR-U model allows for characterizing small-scale fading behavior in a wide range of propagation conditions, ranging from no fading ($K\rightarrow\infty$, $\Delta=0$) to a fading more severe than Rayleigh ($K\rightarrow\infty$, $\Delta\approx1$). The hyper-Rayleigh behavior exhibited by the GTR-U fading model when the two LOS components have equal power (i.e., ${\Delta=1}$) has an intuitive explanation. When $\alpha$ is uniformly distributed, there is a finite probability that $\alpha$ takes values close to $\pi$, i.e. the LOS components are out of phase and are cancelled. This is especially important in the simple Two-Ray (or Two Wave) model, in which the diffuse part is absent; therefore, even in the presence of two very strong LOS components the actual fading behavior is more severe compared to other NLOS models like Rayleigh.

The original consideration of $\alpha$ as uniformly distributed in \cite{durgin2000} is based on two assumptions related to the phase distribution of the waves: (a) uniform distribution for the phase of each one of the individual waves, and (b) uncorrelated phases for each pair of waves. The first assumption is easily justified based on the propagation of electromagnetic waves for distances much larger than their wavelengths. The second assumption comes from the consideration of differently scattered waves. Clearly, this second assumption is well-justified for the diffuse component, but may not hold for some scenarios where the two LOS components are affected by similar scattering. Specifically, in \cite{hyperR1} it was observed that the uniform phase assumption for $\alpha$ may not hold in some practical scenarios where the range of valid angles is limited.

For this reason, a truncated uniform phase model over $[0,1.05\pi)$ was proposed in \cite{hyperR1} to better describe the propagation conditions inside an airframe, where reflections from an enclosed and metallic structure may lead to similar scattering conditions for the two LOS components. Limiting the range of valid phases for $\alpha$ in \cite{hyperR1} caused a worse fading condition than the Two-Ray model. Based on this observation, we note that in the limit case where the phase difference is deterministic $\alpha\equiv\pi$ we would have total cancellation. Hence, using a distribution for $\alpha$ that concentrates the probability close to $\pi$ would cause a more severe fading than in the Two-Ray model. 

A plausible mechanism for such behavior can arise from a grazing reflection on a flat surface; if the angle of incidence is small and the reflection does not cause attenuation, then the two LOS components are likely to have similar amplitudes (i.e. $\Delta \approx 1$) and their phases may also be similar. This effect can be amplified if very directional antennas are used: waves arriving at very specific directions dominate over the diffuse part, and narrow beams cause the range of angles of incidence to be shrunk (or even truncated).

\subsection{Proposed models.}

Our motivation for presenting a new family of fading models is to analyze fading behavior where the phase difference between LOS components is other than uniform, i.e. the range of valid phases for the two LOS components is limited and exhibits some kind of correlation. With the formulation considered in this paper, we now analyze the GTR fading model assuming a more general distribution for the phase difference $\alpha$. For an arbitrarily distributed $\alpha$, from (\ref{eq:condPDF}) we define the envelope pdf of a Generalized Two-Ray fading model as the average of $f_{\text{Rice}}$ over the distribution of $\alpha$. 
\begin{align}
f_{\text{GTR}}(r)&=\int_{0}^{2\pi}f_{\text{Rice}}\left(r;\, K[1+\Delta\cos(\alpha)]\right)f_{\alpha}(\alpha)d\alpha\label{eq:TWDPconditionalFullRice}
\end{align}
Although a different distribution for the GTR fading model arises for any particular choice of $f_{\alpha}(\alpha)$, we will focus on some specific distributions that can help us model harsh propagation conditions.

\begin{figure}[ht]
\includegraphics[width=0.97\columnwidth]{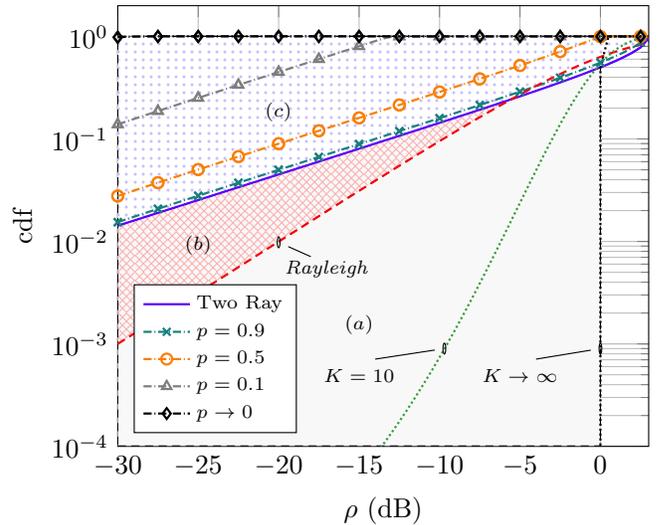}
\caption{The cdf of the GTR-T fading model vs. the normalized envelope amplitude, for different values of the phase truncation parameter $p$. Parameter values for GTR-T fading are $\Delta=1$ and $K\rightarrow\infty$ as in the Two Ray model (obtained for $p=1$), and $\phi=0$. Shaded regions correspond to $(a)$ Rician fading, $(b)$ Hyper-Rayleigh fading, $(c)$ Hyper Two-Ray fading.}
\label{fig1a}
\end{figure}

Following this reasoning, we first study GTR fading with truncated phase (GTR-T), where $\alpha\sim\mathcal{U}(\pi(1-p),\pi(1+p))+\phi$, where $p\in(0,1]$ and $\phi\in(-\pi,\pi)$. This includes the phase model considered in \cite{hyperR1} as a particular case. The envelope pdf in this scenario is given by
\begin{align}
f_{\text{GTR-T}}(r)&=\frac{1}{2\pi p}\int_{\pi(1-p)+\phi}^{\pi(1+p)+\phi}f_{\text{Rice}}\left(r;\, K[1+\Delta\cos(\alpha)]\right)d\alpha\label{eq:GTR-TP}
\end{align}
When $p=1$ and $\phi=0$, the GTR-T fading model reduces to the conventional GTR-U fading model. As $p\rightarrow0$ we observe that the phase is concentrated towards $\pi+\phi$; thus, for $\phi=0$ the probability that the two LOS components cancel each other is increased, causing a fading worse than the conventional Two-Ray fading. Hence, the GTR-T fading distribution can model fading right from extremely favorable propagation conditions when $\Delta=0$ and $K\rightarrow\infty$ to very severe fading close to complete cancellation as $p\rightarrow0$ with $\Delta=1$. This is illustrated in Fig. \ref{fig1a}, where the cdf of the GTR-T fading model is represented and compared with Rician, Rayleigh and Two-Ray fading models. We assume a fixed transmit power $P=1$, and the envelope amplitude is normalized to $\sqrt{\bar\gamma}$ according to eq. (\ref{eq:gamma}).

The GTR-T fading model has an analytically simple formulation since the pdf in (\ref{eq:GTR-TP}) has the same integrand as the conventional GTR fading model. However, it may be argued that a truncated model for the phase $\alpha$ might not be realistic in other situations. For this reason, we present another alternative for the family of GTR fading models.

Now, let us consider that $\alpha$ is distributed according to the von Mises (VM) distribution \cite{vonMises} with pdf given by
\begin{equation}
\label{pdfVM}
f_{\alpha}^{\text{VM}}(\alpha)=\frac{\exp{\left(\eta \cos (\alpha-\varphi)\right)}}{2\pi I_0(\eta)},\,\,\,\,\,\,\,\alpha\in[0,2\pi],
\end{equation}
where $\eta\geq0$ and $\varphi\in\mathbb{R}$ are usually referred to as concentration and centrality parameters. This distribution, also known in the literature as the circular normal distribution or Tikhonov distribution, is widely used in different applications in communications (see \cite{Abreu08} and references therein) to describe the statistics of angles of arrival in wireless systems, or phase error in phased-locked loops (PLLs) just to name a few examples. This model also includes the uniform phase as a particular case when $\eta=0$.

The parameters $\eta$ and $\varphi$ have a similar interpretation as their counterparts $p$ and $\phi$ in the truncated uniform phase model. In both cases $\varphi$ and $\phi$ play the role of centrality parameters and have the same interpretation; however, while both $p$ and $\eta$ concentrate the phase on a given range, they do it in very different ways. One of the potential benefits of using the GTR-V model is that the parameter $\eta$ allows for a smoother transition to the uniform distribution, and does not restrict the range of valid phases as exclusively as $p$ does in the GTR-T model. An additional reason that can justify using the GTR-V model is related to the fact that we are trying to model a phase correlation; this means that these phases influence each other. Since the phase difference between two correlated sources is modeled by the von Mises distribution \cite{Abreu08}, it makes sense to use this distribution for building a generalized two-ray model in this context. Lastly, an additional advantage of the GTR-V model over the GTR-T model is purely mathematical, since as we will later see the MGF of the GTR-V can be computed in closed-form. Hence, this may be useful to analyze system performance metrics in a simpler way.

Since we are interested in modeling hyper-Rayleigh behavior, the centrality parameter is set to $\varphi=\pi$. Thus, for $\eta\neq 0$ the probability of $\alpha$ taking values close to $\pi$ increases as $\eta$ increases. With this consideration, the pdf for the GTR fading model with VM-distributed $\alpha$ (GTR-V) is given as
\begin{align}
\label{eq:GTR-VM}
&f_{\text{GTR-V}}(r)=\frac{1}{2\pi I_0(\eta)}\times\\&\int_{0}^{2\pi}f_{\text{Rice}}\left(r;\, K[1+\Delta\cos(\alpha)]\right)\exp{\left(-\eta \cos \alpha\right)}d\alpha\nonumber
\end{align}

\begin{figure}
\includegraphics[width=0.97\columnwidth]{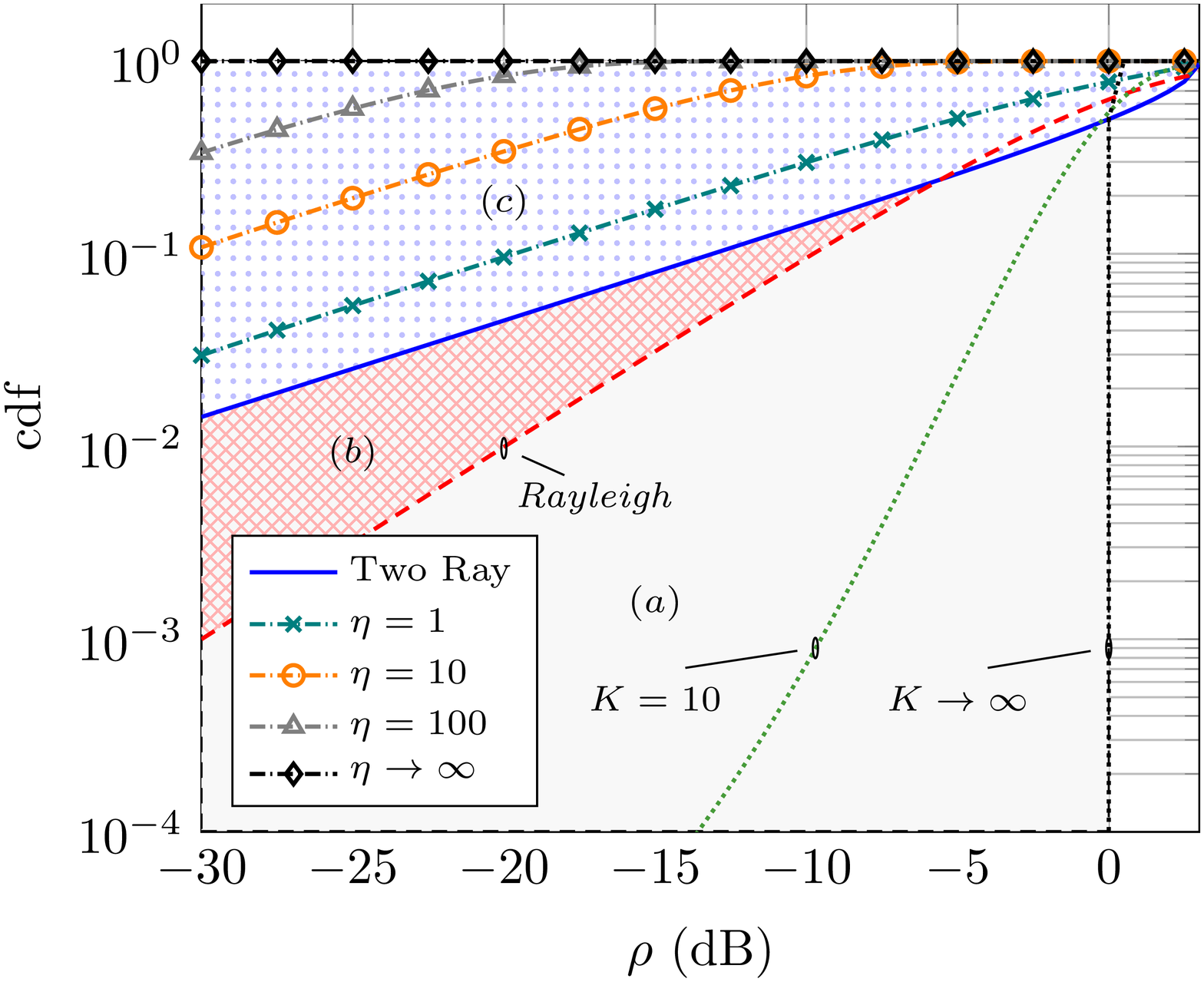}
\caption{The cdf of the GTR-V fading model vs. the normalized envelope amplitude, for different values of the phase scale parameter $\eta$. Parameter values for GTR-V fading are $\Delta=1$ and $K\rightarrow\infty$ as in the Two-Ray model (obtained for $\eta=0$). Shaded regions correspond to $(a)$ Rician fading, $(b)$ Hyper-Rayleigh fading, $(c)$ Hyper Two-Ray fading.}
\label{fig1}
\end{figure}

The behavior of the GTR-V fading model is shown in Fig. \ref{fig1}. We observe that as $\eta$ grows and $\Delta=1$, the fading falls in the region beyond the Two Ray model; hence, it is also suitable for characterizing very severe propagation conditions\footnote{In the literature, there are other fading models that can be used for modeling hyper-Rayleigh behavior. Specifically, the well-known Nakagami-$m$ model for $m\in[0.5,1)$ is useful to model hyper-Rayleigh behavior, being coincident with the Rayleigh fading model for $m = 1$. However, the Nakagami-$m$ model is only linked to a physically-justified model when $m$ is an integer or a half-integer. Therefore the physical interpretation of such a model is lost as opposed to GTR models.
The $\kappa$-$\mu$ fading model \cite{kunumodel} was proposed as a generalization of Rician fading model, and it can be useful to model hyper-Rayleigh behavior when the parameter $\kappa$ (associated with the ratio between LOS and NLOS power similarly to Rician $K$ parameter) tends to infinity, and the parameter $\mu$ (associated to the number of multipath clusters that form the diffuse component) tends to zero, but their product $\kappa\cdot \mu$ remains constant. This special regime of the $\kappa$-$\mu$ model was denoted as Extreme $\kappa$-$\mu$ fading \cite{Extreme}, and is also valid for modeling hyper-Rayleigh behavior in a different way as the GTR models (the slope of the cdf of both models is different). However, as discussed in \cite{Extreme}, in this limiting regime this model loses its physical interpretation for the set of values from which it is built.
For the readers' convenience, a graphical comparison between these other families of fading models in the context of hyper-Rayleigh fading similar to Figs. \ref{fig1a} and \ref{fig1} can be found in \cite{WorseThanRay,hyperR1,Extreme}.}.

\subsection{Effect on the average SNR at the receiver.}
The distributions in (\ref{eq:GTR-TP}) and (\ref{eq:GTR-VM}) can model the effect of a larger cancellation of the LOS components due to the statistical behavior of the phase difference $\alpha$. However, we also note that by simply applying a deterministic shift of value $\pi$ to this phase difference, the resulting distributions would be centered in zero. This implies that the two LOS components would be cancelled with less probability, and hence the fading experienced by the signal would be closer to a Rician behavior rather than to a hyper-Rayleigh behavior. This can be seen by deriving the expression for the average SNR of these models using Lemma \ref{lemma1}:
\begin{align}
\mathbb{E}_{\text{GTR-T}}(\gamma)=&\bar{\gamma}\left(1\mp\Delta\frac{K}{K+1}\text{sinc}(p)\right),\\
\mathbb{E}_{\text{GTR-V}}(\gamma)=&\bar{\gamma}\left(1\mp\Delta\frac{K}{K+1}\frac{I_1(\eta)}{I_0(\eta)}\right),
\end{align}
where the function $\text{sinc}(p)=\sin(\pi p)/\pi p$ and $\bar\gamma$ is the average SNR of the conventional GTR fading model. In these expressions, the negative sign accounts for the cases where the distribution of $\alpha$ is centered at $\pi$, whereas the positive sign corresponds to the case where the distributions are centered at zero.

It is interesting to observe how the average SNR is reduced in three circumstances: $(1)$ when the two LOS components tend to have similar magnitudes (i.e. increasing $\Delta$), $(2)$ when the LOS power is larger (i.e. increasing $K$) and $(3)$ when the phase $\alpha$ is more concentrated towards $\pi$ (i.e. reducing $p$ or increasing $\eta$). It is easy to see how in the limiting cases of the three parameters (i.e. $\Delta\rightarrow1$, $K\rightarrow\infty$ and $p\rightarrow 0$ or $\eta\rightarrow\infty$), the average SNR tends to zero. 

\begin{table*}
\caption{MGF-based calculation of SEP. Single-channel and multi-channel reception with different detection strategies: $M$-ary Phase-Shift Keying (M-PSK), Quadrature Amplitude Modulation (M-QAM), Differential PSK (M-DPSK), and Frequency-Shift Keying (M-FSK), cfr. \cite[eq. 9.15, 9.21, 8.200, 8.192]{AlouiniBook}}.
\vspace{-5mm}
\begin{center}
  \renewcommand{\arraystretch}{2}
\begin{tabular}[c]{ |l|l| }
%
  \hline
  \multicolumn{2}{|c|}{\textsc{Single channel reception, Coherent detection}} \\[1pt]
  \hline
  \hline
  M-PSK & $P_S(\bar\gamma)= \I_{(M-1)\pi/M}\left[\M\left(-\frac{\sin^2 \pi/M}{\sin^2\theta},K,\Delta,\bar\gamma\right)\right]$ \\
  \hline
  M-QAM & $P_S(\bar\gamma)= 4c\I_{\pi/2}\left[\M\left(-\frac{3}{2(M-1)\sin^2\theta},K,\Delta,\bar\gamma\right)\right]-4c^2\I_{\pi/4}\left[\M\left(-\frac{3}{2(M-1)\sin^2\theta},K,\Delta,\bar\gamma\right)\right];\,\,\,\,c=\left(1-\frac{1}{\sqrt{M}}\right)$ \\
  \hline\hline
  \multicolumn{2}{|c|}{\textsc{Single channel reception, Differentially-coherent detection}} \\[1pt]
  \hline
  \hline
   M-DPSK & $P_S(\bar\gamma)= \I_{(M-1)\pi/M}\left[\M\left(-\frac{\sin^2 \pi/M}{1+\cos{\pi/M}\cos\theta},K,\Delta,\bar\gamma\right)\right]$ \\
  \hline\hline
   \multicolumn{2}{|c|}{\textsc{Single channel reception, Non-coherent detection}} \\[1pt]
  \hline
  \hline
   M-FSK & $P_S(\bar\gamma)= \sum_{m=1}^{M-1}(-1)^{m+1}\left(\bfrac{M-1}{m}\right)\frac{1}{m+1}\M\left(-\frac{m}{m+1},K,\Delta,\bar\gamma\right)$ \\
     \hline
  \hline
  \multicolumn{2}{|c|}{\textsc{Multichannel reception, Coherent detection}} \\[1pt]
  \hline
  \hline
  M-PSK & $P_S(\bar\gamma)= \I_{(M-1)\pi/M}\left[\prod_{i=1}^{L}\M\left(-\frac{\sin^2 \pi/M}{\sin^2\theta},K_i,\Delta_i,\bar\gamma_i\right)\right]$ \\
  \hline
  M-QAM & $P_S(\bar\gamma)= 4c\I_{\pi/2}\left[\prod_{i=1}^{L}\M\left(-\frac{3}{2(M-1)\sin^2\theta},K_i,\Delta_i,\bar\gamma_i\right)\right]-4c^2\I_{\pi/4}\left[\prod_{i=1}^{L}\M\left(-\frac{3}{2(M-1)\sin^2\theta},K_i,\Delta_i,\bar\gamma_i\right)\right]$ \\
  \hline
\end{tabular}
\label{table1}
\end{center}
\end{table*}

However, by simply concentrating the phase $\alpha$ towards zero, we cause the average SNR to be increased by the same magnitude. In the limiting case previously discussed, we would be increasing the average SNR by a factor of 2. We must note that the standard comparison between fading models in the literature is usually tackled by having them normalized to their average SNR. However, we wanted to focus on a different aspect: the effect of having a correlated phase difference on the average SNR. Therefore, for a fixed transmit power $P$, we could have a larger or smaller average SNR at the receiver due to phase correlation. In that sense, the investigated models can actually be more detrimental than the conventional GTR model (for a fixed transmit power); equivalently, systems operating in these scenarios would require more transmit power to attain a target average SNR. Intuitively, when we consider two LOS components there is a large amount of transmit power that may or may not be collected at the receiver side depending on behavior of the relative phase of these LOS components.

It is also easy to see how the MGF for the GTR-V fading model can be conveniently expressed in closed-form. Using Lemma \ref{lemma1} including the pdf of the VM distribution in (\ref{pdfVM}) and (\ref{eq:mgfRice}), we directly obtain:
\begin{align}
&\mathcal{M}_{\text{GTR-V}}(s)=\nonumber\\
&\frac{1+K}{1+K-s\bar{\gamma}}\exp\left(\frac{Ks\bar{\gamma}}{1+K-s\bar{\gamma}}\right)\frac{I_{0}\left(\pm\eta-\frac{Ks\bar{\gamma}\Delta}{1+K-s\bar{\gamma}}\right)}{I_0(\eta)},\label{eq:MTWDPVM}
\end{align}
where the $\pm$ signs correspond to the distribution of $\alpha$ concentrated towards $\pi$ and zero, respectively.

\section{System Performance Analysis Using the MGF}
\label{MGFSysPer}
We have presented a connection between the Rician fading model and a class of Generalized Two-Ray models, that has allowed us to obtain a simple expression for the MGF. This allows us to easily conduct the analysis of different system performance metrics. Specifically, we use our results to analyze the SEP and the ergodic capacity in GTR-U fading channels, as presented below. 

\subsection{Symbol Error Probability}
The average probability of symbol error $P_{S}$ of a fading channel is given by \cite{Goldtext} as
\begin{equation}
P_{S}(\bar{\gamma})=\int_{0}^{\infty}P_{AWGN}(\gamma)f_{\gamma}(\gamma)d\gamma,
\label{eq:Perror}
\end{equation}
where $P_{AWGN}(\gamma)$ is the probability of symbol error of an AWGN channel with SNR $\gamma$. Using the MGF approach \cite{AlouiniBook}, the resultant expression for the SEP is an integral of a smooth finite integrand over finite limits; in some cases, the SEP is given directly in terms of the MGF. These expressions are summarized in Table \ref{table1} where we have defined the auxiliary function
\begin{equation}
\label{eq:auxFun}
\I_{\beta}\left[f(\theta)\right]\triangleq \frac{1}{\pi}\int_{0}^{\beta}f(\theta)d\theta
\end{equation}
for the sake of compactness. For the specific case of multichannel reception with $L$ independent branches using Maximal Ratio Combining (MRC), the average SNR at the receiver is given by $\bar\gamma=\sum_{i=1}^{L}\bar\gamma_i$, and the MGF of interest is expressed as the product of the individual MGFs per receive branch.

As opposed to existing analyses in the literature, we note that the expressions in Table \ref{table1} are exact, and allow for characterizing the SEP in some of the scenarios considered in \cite{BPSKTWDP,MRCTWDP,QAMTWDP} following a unified approach.

Of special interest is the particular case of the binary DPSK modulation scheme, where the SEP (equivalent to the BER) can be obtained in closed-form as
\begin{align}
P_{S}(\bar{\gamma})&=\frac{1}{2}\mathcal{M}_{\text{GTR-U}}(-1;\, K,\Delta)\nonumber\\
&=\tfrac{1}{2}\tfrac{1+K}{1+K+\bar{\gamma}}\exp\left(\tfrac{-K\bar{\gamma}}{1+K+\bar{\gamma}}\right)I_{0}\left(\tfrac{K\bar{\gamma}\Delta}{1+K+\bar{\gamma}}\right).\label{eq:PSS}
\end{align}
This simple case can provide important insights about the effect of the parameter $\Delta$ on the error probability. We observe that the BER under the GTR fading model can be seen as the BER under the Rician case modulated by a term that depends on the modified Bessel function $I_0(\cdot)$. The Bessel function term is always greater than one except for the case when $\Delta=0$; hence, being a monotonically increasing function, the error increases as $\Delta$ increases. Specifically, the hyper-Rayleigh behavior exhibited by GTR-U fading model when $\Delta\approx1$ and $K\rightarrow\infty$ leads to the BER in this scenario to be expressed as
\begin{align}
\label{eq:SEP}
P_{S}(\bar{\gamma})|_{K\rightarrow\infty}&\approx\frac{1}{2}\exp\left(-\bar\gamma\right)I_{0}\left(\Delta\bar\gamma\right).
\end{align}

\begin{table*}
\caption{Asymptotic results for the ergodic Capacity in GTR-U fading (perfect CSI at the receiver) in the high-SNR regime.}
\vspace{-5mm}
\begin{center}
  \renewcommand{\arraystretch}{2}
\begin{tabular}[c]{ |l|l| }
  \hline
  Rice & $C_{\text{ora}}|_{\bar\gamma\Uparrow}\approx \nu\cdot\bar\gamma(dB) + \log_2 e\left\{\log \left(\frac{K}{K+1}\right)+\Gamma(0,K)\right\}$ \\
  \hline
  GTR-U & $C_{\text{ora}}|_{\bar\gamma\Uparrow}\approx  \nu\cdot\bar\gamma(dB)+\log_2 e \left\{\log \left(\frac{K}{K+1}\right)+\log\left(\frac{1+\sqrt{1-\Delta^2}}{2}\right)+\mathcal{J}(K,\Delta)\right\},\,\,\,\,\,\,\,\,\,\,\mathcal{J}(K,\Delta)=\int_1^{\infty} \frac{e^{-tK}}{t}I_0(tK\Delta) dt.$ \\
 \hline
  GTR-U${}_{(K\cdot\Delta>>1)}$& $C_{\text{ora}}|_{\bar\gamma\Uparrow}\approx  \nu\cdot\bar\gamma(dB)+\log_2 e \left\{ \log \left(\frac{K}{K+1}\right)+\log\left(\frac{1+\sqrt{1-\Delta^2}}{2}\right)+\sqrt{\frac{2}{\pi}}\left[\frac{e^{-K(1-\Delta)}}{\sqrt{K\Delta}}-\sqrt{\left(\tfrac{1}{\Delta}-1\right)}\text{erfc}\left(K(1-\Delta)\right)\right]\right\}$ \\
 \hline
    GTR-U${}_{(K>>1,\Delta=1)}$ & $C_{\text{ora}}|_{\bar\gamma\Uparrow}\approx  \nu\cdot\bar\gamma(dB)+ \log_2 e \left\{ \log \left(\frac{K}{K+1}\right)-\log2+\sqrt{\frac{2}{\pi K}}\right\}$ \\ 
 \hline
  $C$ loss GTR-U & $\delta_C(K,\Delta)= \log_2 e \left\{\Gamma(0,K)-\log\left(\tfrac{1+\sqrt{1-\Delta^2}}{2}\right)-\mathcal{J}(K,\Delta)\right\}$ \\   
 \hline
  $C$ loss Two-Ray & $\delta_C(K\rightarrow\infty,1)= 1$ \\   
   \hline
\end{tabular}
\label{table2}
\end{center}
\end{table*}

We see that for $\Delta=0$ the error probability reduces to the AWGN case, as no fading occurs for $K\rightarrow\infty$; however, we see that the impact of $\Delta>0$ is captured by the fact that $I_0(\Delta\bar\gamma)>1$. 

When $\bar\gamma\to\infty$ in (\ref{eq:PSS}), we have that the asymptotic BER is given by
\begin{align}
\label{eq:SEPasy}
P_{S}|_{\bar{\gamma\rightarrow\infty}}&\approx\frac{1}{2}\frac{1}{1+\bar\gamma/(K+1)}\exp\left(-K\right)I_{0}\left(\Delta K\right).
\end{align}
Again, the effect of having two LOS components appears in the argument of the Bessel function $I_{0}\left(\Delta K\right)>1$. 

\subsection{Ergodic Capacity}
\label{sec:C}
The effect of fading on the maximum rate of data transmission over a wireless link has been a matter of interest in communication and information theory for many years, considering different adaptation policies at the transmitter and receiver sides, as well as for different configurations in terms of the number of antennas. Specifically, the work by Alouini and Goldsmith \cite{Alouini1999} provided the first analytical results for the capacity of adaptive transmission with diversity-combining techniques in Rayleigh fading. However, extensions of these results to other types of fading are often more challenging and do not lend themselves to analytically tractable solutions.

Inspired by the general framework for the average error probability analysis based on the MGF \cite{AlouiniBook}, an alternative formulation for the analysis of the ergodic capacity in fading channels in terms of the MGF of the received SNR was recently proposed in \cite{DiRenzo}, and was then further complemented in \cite{yilmaz2012}. If the MGF of interest has an analytical closed-form solution, the capacity can be evaluated using a single integral over the MGF.

As an application of this method for evaluating the Shannon capacity in fading channels, we will consider an optimal rate adaptation (ORA) policy with constant transmit power. This is the capacity of the fading channel when the channel state information is only available at the receiver side. According to \cite[eq. 7]{DiRenzo}, the capacity per unit bandwidth is given in terms of the MGF of the SNR at the receiver side as
\begin{align}
C_{\text{ora}}=\log_2e \int_{0}^{\infty}{E_i(-s)\M_{\gamma}^{(1)}(-s)ds},
\label{eq:Cora}
\end{align}
where $E_i(\cdot)$ denotes the Exponential integral function \cite[eq. 2.325.1]{Gradstein2007} and $\M_{\gamma}^{(1)}(s)$ indicates the first derivative of the MGF with respect to $s$. Assuming a multiantenna receiver with $L$ independent branches using MRC detection, we have
\begin{align}
\label{eq:MGFder}
\M_{\gamma}^{(1)}(s)=\sum_{l=1}^{L}\M_{\gamma_l}^{(1)}(s)\times\prod_{\bfrac{k=1}{k\neq l}}^L{\M_{\gamma_k}(s)}.
\end{align}
Since we have a closed-form expression for the MGF of the received SNR per branch for the GTR-U fading model, we can also compute its first derivative in closed-form as
\begin{align}
\label{eq:dMGF}
&\M_{\gamma_l}^{(1)}(s)=\frac{(1+K_l)\bar{\gamma_l}}{(1+K_l-s\bar{\gamma_l})^{2}}\exp\bigg(\frac{K_l s\bar{\gamma_l}}{1+K_l-s\bar{\gamma_l}}\bigg)\times\\&\left[I_{0}\left(\tfrac{K_l s\bar{\gamma_l}\Delta_l}{1+K_l-s\bar{\gamma_l}}\right)\left(1+\tfrac{K_l(1+K_l)}{1+K_l-s\bar{\gamma_l}}\right)+\tfrac{K_l\Delta(1+K_l)}{1+K_l-s\bar{\gamma_l}}I_{1}\left(\tfrac{K_ls\bar{\gamma_l}\Delta_l}{1+K_l-s\bar{\gamma_l}}\right)\right]\nonumber
\end{align}
where $I_1(\cdot)$ is the modified Bessel function of the first kind and order one. Hence, the expression for the ergodic capacity in GTR fading channels using ORA policy and MRC detection can be computed by plugging (\ref{eq:dMGF}) and (\ref{eq:MGFder}) into (\ref{eq:Cora}).

Using \cite[eq. 12]{DiRenzo}, we find a simple asymptotic approximation for the capacity in the low-SNR regime as
\begin{align}
\label{eq:Clow}
C_{\text{ora}}|_{\bar\gamma\Downarrow}\approx\log_2e{\M_{\gamma}^{(1)}(s)}|_{s=0}=\bar\gamma\log_2e,
\end{align}
where we assumed that the received SNRs per branch are i.i.d. and $\bar\gamma=L\bar\gamma_l$. Interestingly, we observe that (\ref{eq:Clow}) is independent of $\Delta$ for GTR fading.

An asymptotic expression for the capacity in the high-SNR\footnote{Note that at high-SNR, the capacity with ORA policy is the same as the capacity with optimal power and rate allocation (OPRA) policy, which considers that CSI is available at both the transmitter and receiver sides \cite{Alouini1999}.} can also be obtained from the first derivative of the $n^{th}$ moment \cite[eq. 8]{Yilmaz2012b} or \cite[eq. 22]{ansari2014} as
\begin{align}
\label{eq:Chigh}
C_{\text{ora}}|_{\bar\gamma\Uparrow}\approx \log_2 e \cdot \frac{\partial}{\partial n} \mathbb{E}\left[\gamma^n\right]|_{n=0}.
\end{align}

In Table \ref{table2}, we summarize the asymptotic results (high-SNR) for the capacity in GTR fading and a single-branch receiver. In Appendix \ref{app1}, we first obtain $C_{\text{ora}}|_{\bar\gamma\Uparrow}$ for a Rician fading channel. The derivations for the GTR case are included in Appendix \ref{app2}, where the asymptotic capacity is given in the form $C_{\text{ora}}|_{\bar\gamma\Uparrow}=\nu \cdot \bar\gamma (dB) + \mu$, where $\nu=0.1\log(10)\log_2(e)$, $\mu$ is a constant value independent of the average SNR, and the average SNR $\bar\gamma (dB) = 10\log_{10}\bar\gamma$ is given in dB.

The capacity loss or the difference between the asymptotic capacity of Rice and GTR, given by $\delta_C=C_{\text{ora}}^{\text{Rice}}-C_{\text{ora}}^{\text{GTR-U}}$ is,
\begin{align}
\label{eq:closs1}
\delta_C(K,\Delta)=\log_2 e \left\{\Gamma(0,K)-\log\left(\tfrac{1+\sqrt{1-\Delta^2}}{2}\right)-\mathcal{J}(K,\Delta)\right\}.
\end{align}
It is easy to verify that $\delta_C>0$. In the hyper-Rayleigh zone of the GTR fading model, we have that the capacity loss is
\begin{align}
\label{eq:closs2}
\delta_C(K\rightarrow\infty,\Delta=1)=1
\end{align}
with respect to the AWGN case (i.e. Rician with $K\rightarrow\infty)$. This implies that the capacity loss in the most severe fading condition modeled by GTR fading is only $1$ bps/Hz with respect to the AWGN case (i.e. no fading).

\section{System Design Implications}
\label{SysTradeOff}
The preceding analysis allows us to gain new insights on the behavior of the GTR fading model. In this section, we evaluate the derived expressions for the system performance metrics of GTR fading in some scenarios of interest. As in the previous section, we well consider that $\alpha$ is uniformly distributed in $[0,2\pi]$.

\begin{figure}
\includegraphics[width=0.97\columnwidth]{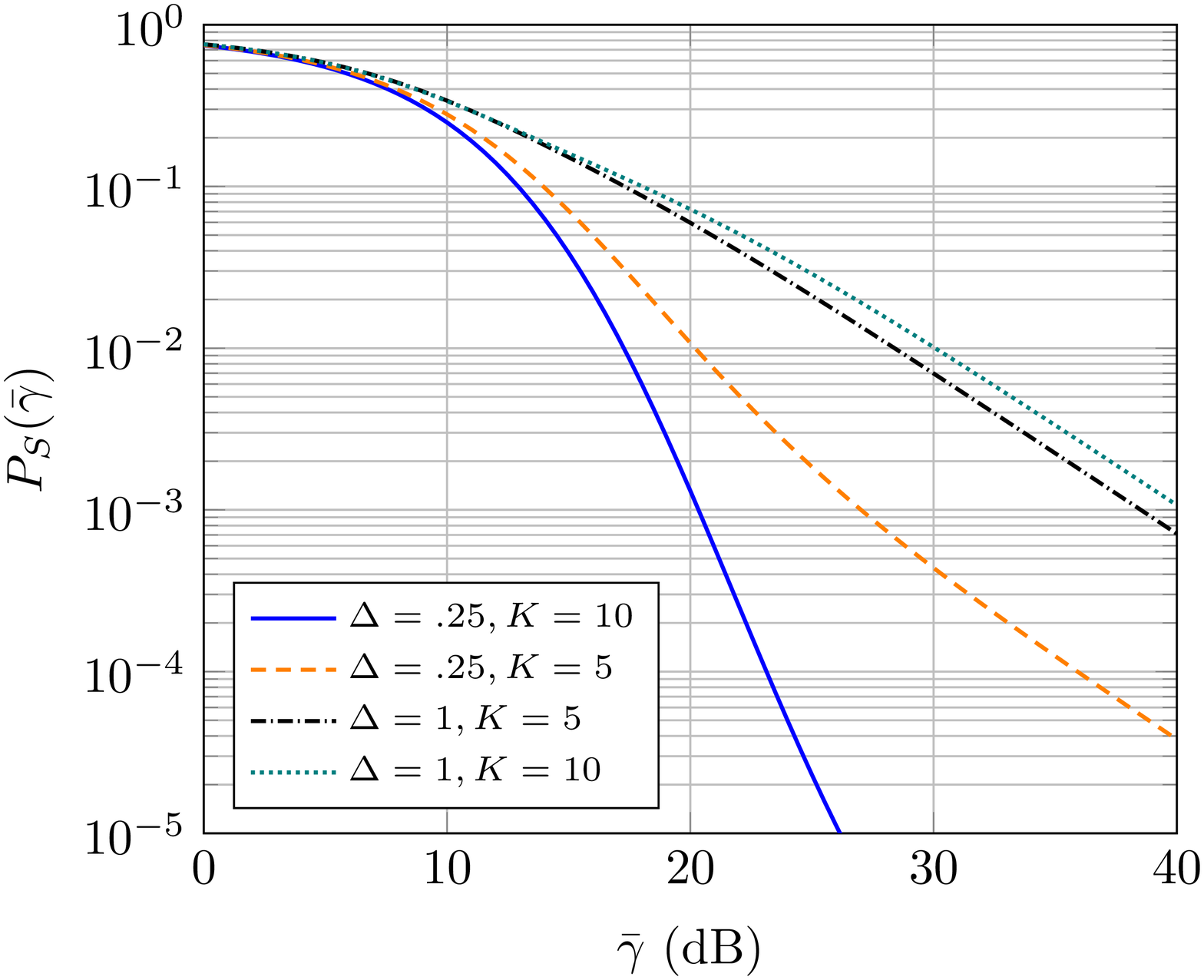}
\caption{SEP vs average symbol SNR $\bar\gamma$, for $16$-QAM modulation scheme and different fading conditions. Single-branch reception with coherent detection is considered.}
\label{fig:SEP01}
\end{figure}

We first evaluate the symbol error probability in GTR fading channels considering fading conditions and numbers of receive antennas. In Fig. \ref{fig:SEP01}, we illustrate the effect of the parameters $K$ and $\Delta$ on the SEP of $16$-QAM modulation scheme with coherent detection. Similar conclusions can be obtained for the remaining modulation and detection schemes summarized in Table \ref{table1}. When the value of $\Delta$ is low, we observe that the error probability is reduced as $K$ grows; conversely, in the limit case of $\Delta=1$ we notice that the error probability is increased for larger values of $K$. This is consistent with the observation made in equation (\ref{eq:SEP}); as $K\rightarrow\infty$ the fading becomes more severe if $\Delta=1$.

\begin{figure}
\includegraphics[width=0.97\columnwidth]{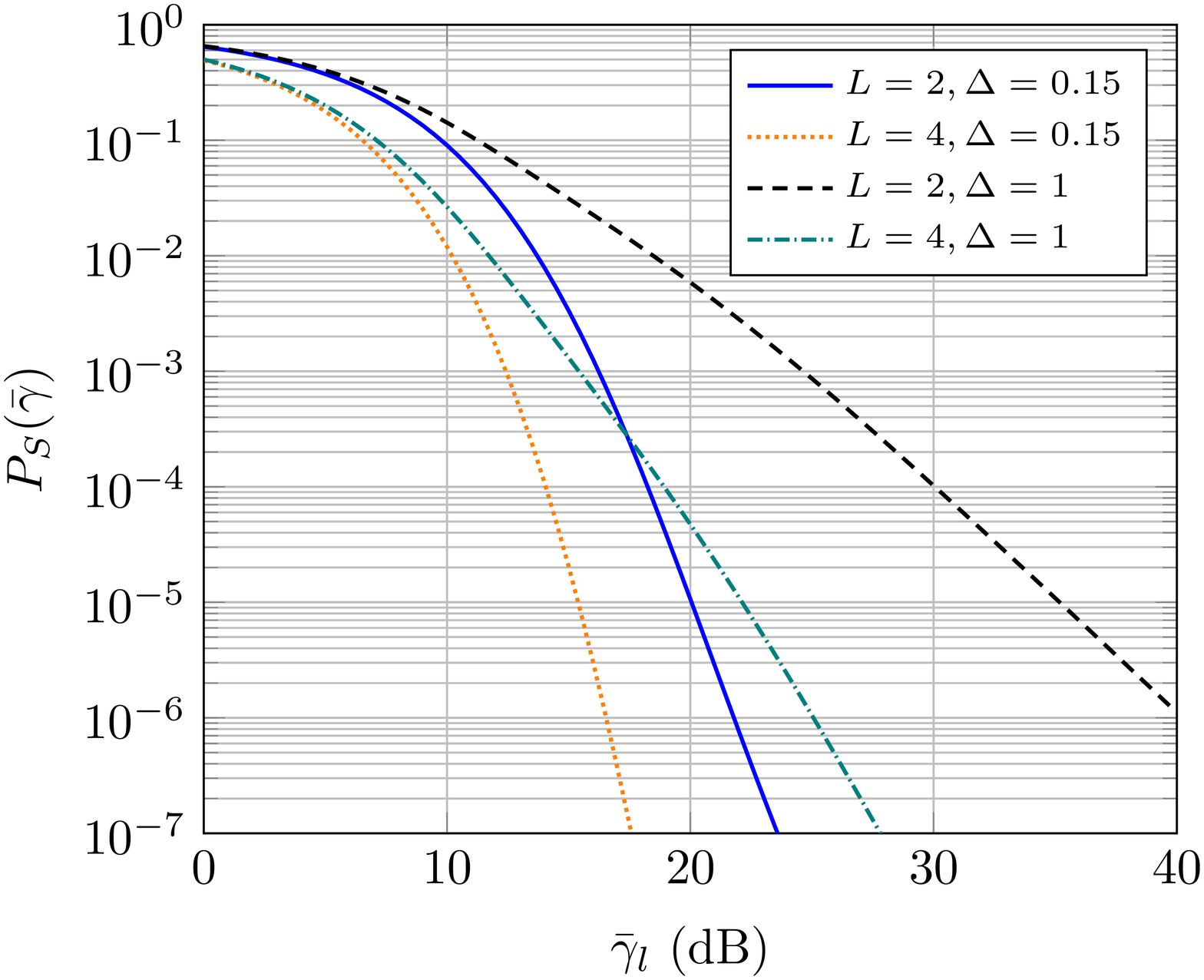}
\caption{SEP vs average SNR per branch $\bar\gamma_l$, for $16$-QAM modulation scheme and different fading conditions and number of receive branches $L$. Parameter values are $K=10$, $M=16$. Blue: $\Delta=0.15$, Black: $\Delta=1$.}
\label{fig:SEP02}
\end{figure}

In Fig. \ref{fig:SEP02} we investigate the effect of using more receive antennas in the SEP, assuming MRC reception with coherent detection. We consider a value of $K=10$. As in the previous case, as $\Delta$ is increased, the performance is degraded. However, we see that using more receive antennas is extremely beneficial when $\Delta=1$, as the SEP is improved more significantly, compared to $\Delta=0.15$. This is in agreement with the results in \cite{bakir2009} for the Two Ray fading model (i.e., $K\rightarrow\infty$ and $\Delta=1$).

\begin{figure}
\includegraphics[width=0.97\columnwidth]{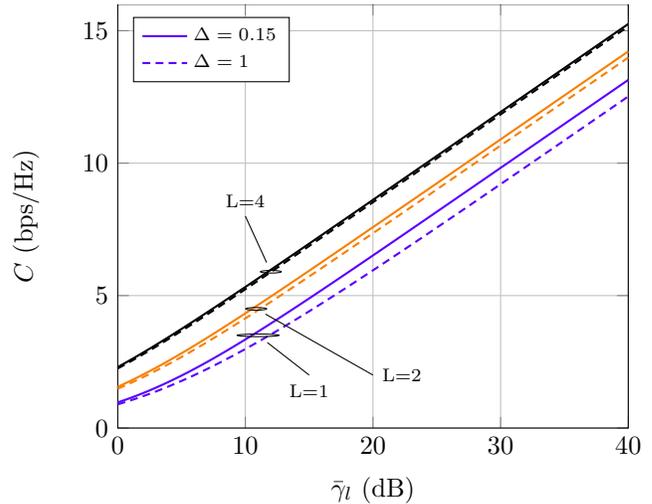}
\caption{Capacity vs average SNR per branch $\bar\gamma_l$, for different fading conditions and numbers of receive antennas $L$. Parameter value $K=10$. Solid lines correspond to $\Delta=0.15$, dashed lines correspond to $\Delta=1$.}
\label{fig:C01}
\end{figure}

We now evaluate the Shannon capacity in GTR fading channels with perfect CSI at the receiver, using (\ref{eq:Cora}) and (\ref{eq:MGFder}). First, we consider an $L$-branch receiver with MRC reception, and we assume a LOS power ratio $K=10$. In Fig. \ref{fig:C01}, we represent the ergodic capacity as a function of the average SNR per branch $\bar\gamma_l$, for different values of the parameter $\Delta$. For the sake of simplicity, we assume i.i.d. receive branches. 

We notice that the capacity is reduced as $\Delta$ grows, leading to a gap for high SNR of around $2$ dB when single antenna reception is used. However, as the number of receive antennas is increased, we see that the capacity is barely affected by the value of $\Delta$. Hence, in very severe fading conditions the use of diversity reception techniques allows for an increase in the capacity.

\begin{figure}
\includegraphics[width=0.97\columnwidth]{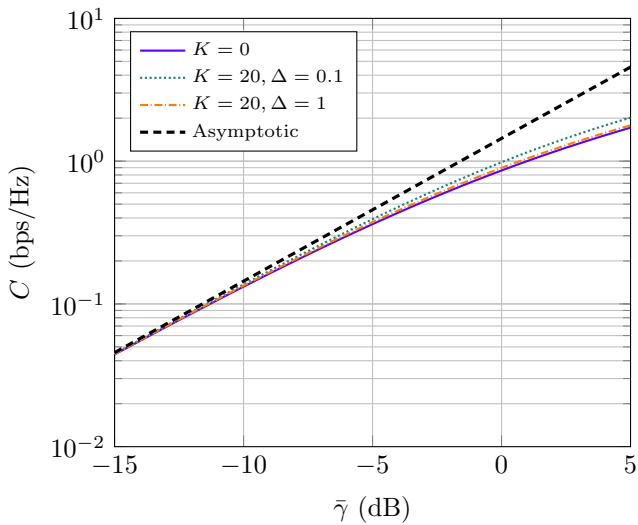}
\caption{Capacity vs average SNR $\bar\gamma$ in the low-SNR regime, for different fading conditions.}
\label{fig:C02}
\end{figure}

We now study the behavior of capacity in the low-SNR and high-SNR regimes. First, in Fig. \ref{fig:C02} we investigate the capacity in the low-SNR regime using the asymptotic approximation given in (\ref{eq:Clow}), as a function of the average SNR $\bar\gamma$ with $L=1$. In the low-SNR regime, we observe that the capacity is asymptotically independent of $K$ and $\Delta$, as suggested by equation (\ref{eq:Clow}).

\begin{figure}
\includegraphics[width=0.97\columnwidth]{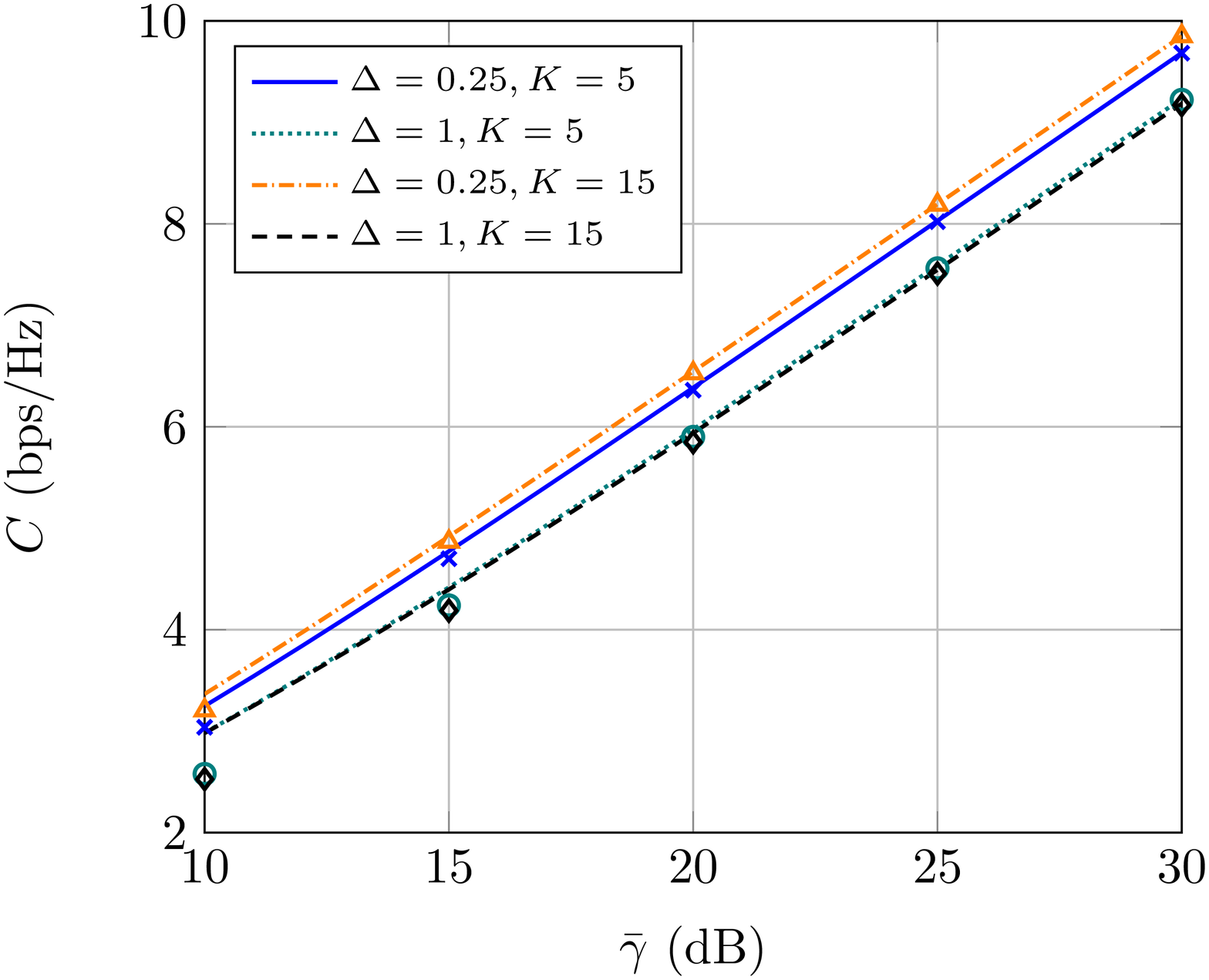}
\caption{Capacity vs average SNR $\bar\gamma$ in the high-SNR regime, for different fading conditions. Markers indicate the asymptotic result given by (\ref{eq:Chigh}).}
\label{fig:C03}
\end{figure}

In Fig. \ref{fig:C03}, the high-SNR regime is considered. The asymptotic capacity results are given by (\ref{eq:Chigh}) and the expressions are summarized in Table \ref{table2}. We see that the asymptotic capacity (represented with markers) is very tight for values of $\bar\gamma>15$ dB and is even more accurate for low values of $\Delta$. Thus, the ergodic capacity in GTR fading channels are well approximated by the simple closed-form expressions derived in Appendix \ref{app2}.

\begin{figure}
\includegraphics[width=0.97\columnwidth]{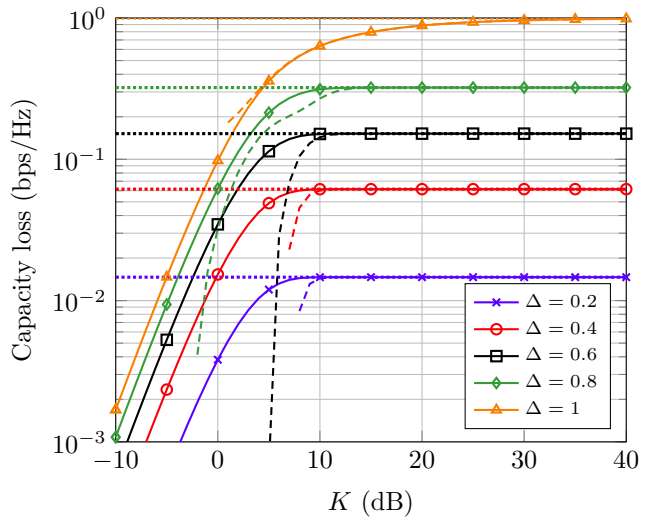}
\caption{Capacity loss in GTR-U fading with respect to Rician fading, as a function of the LOS power factor $K$ (in dB). Solid lines correspond to the exact evaluation of $\mathcal{J}(K,\Delta)$ in (\ref{eq:intapp}), dashed lines correspond to the approximate evaluation of $\mathcal{J}(K,\Delta)$ in (\ref{eq:intapp2}) for large $K\cdot\Delta$, and dotted lines correspond to the asymptotic case of $K\rightarrow\infty$.}
\label{fig:C04}
\end{figure}

Fig. \ref{fig:C04} represents the asymptotic capacity loss of GTR fading channels with respect to the case of Rician fading (i.e., $\Delta=0$). This metric $\delta_C(K\Delta)$ is independent of $\bar\gamma$, and indicates how the capacity is reduced due to the non-zero probability of the two LOS components partially cancelling, dependent on the parameter $\Delta$. We represent this capacity loss as a function of the LOS power ratio parameter $K$, for different values of $\Delta$.

As $K$ is increased, the capacity loss grows to a maximum degradation value given by 
\begin{align}
\delta_C(K\rightarrow\infty,\Delta)=1- \log_2 \left(1+\sqrt{1-\Delta^2}\right)
\end{align}
that corresponds to the capacity reduction with respect to the AWGN case. We see how the approximate expression for $\mathcal{J}(K,\Delta)$ is very accurate for reasonably large values of $K\cdot\Delta$. In the limiting case of the hyper-Rayleigh fading condition (i.e. $K\rightarrow\infty$ and $\Delta=1$), we see that the capacity loss is only $1$ bps/Hz.

\vspace{-3mm}
\section{Conclusion}
\label{conclusion}

We have provided an analytical approach to the characterization of Generalized Two Ray fading channels, and systems operating over them. By observing that the GTR fading conditioned on the difference in phase between the two LOS components results in Rician fading, any linear metric of the GTR fading can be expressed in terms of a simple finite integral of the corresponding metric of the Rice fading model. This simple yet powerful approach has allowed us to derive a closed-form expression for the MGF of the GTR-U fading model for the first time in the literature. We also provided very simple expressions for the relevant performance metrics of systems experiencing GTR-U fading such as the amount of fading and the level crossing rate.  

We then used this unified MGF-based technique to analyze the error rate performance of different modulation schemes and detection techniques. Our results provided interesting insights on the effect of the parameters $K$ and $\Delta$ on the symbol error probability and implications on system design. We also investigated the capacity limits of communication systems affected by GTR-U fading and observed that the asymptotic capacity penalty per bandwidth unit in the extreme case of hyper-Rayleigh fading with respect to the AWGN case is only $1$ bps/Hz in the high-SNR regime, when perfect CSI is available at the receiver.

Leveraging the connection between the Rician distribution and the GTR-U fading model, we have shown that the latter can actually be generalized into a new family of fading models that can characterize a wider set of propagation conditions: the GTR fading models with arbitrary phase. We showed that considering any phase distribution other than uniform for this model has an impact on the received SNR. We showed that the analytical characterization of this new GTR fading model is of similar complexity to the conventional GTR-U fading model. The empirical validation of this new model with field measurements, as well as the performance limits of communication systems operating in such conditions, will be a matter of future work.


\appendices
\section{Asymptotic capacity of Rician fading}
\label{app1}
The $n^{th}$ moment of the SNR in Rician fading is given by
\begin{equation}
\label{eq:app01}
\mathbb{E}(\gamma^{n})=\frac{n!}{(1+K)^{n}}{_{1}}F_{1}(-n,\,1;\,-K)\bar{\gamma}^{n},
\end{equation}
The derivative of (\ref{eq:app01}) with respect to $n$ evaluated at $n=0$ can be computed using the chain rule. First, we calculate the expression for the derivatives of the confluent hypergeometric function with respect to its first parameter \cite[eq. 38]{ancarani}
\begin{align}
\label{eq:app02}
G^{(1)}(0,1,-K)&=-\frac{\partial}{\partial n}\left[{_{1}}F_{1}(-n,\,1;\,-K)\right]|_{n=0}\\&=K\cdot{}_2F_2(1,1;2,2;-K).
\end{align}
Then, we simplify the generalized hypergeometric function for the particular values of its arguments as suggested in \cite[eq. 10]{Lin2005}
\begin{align}
\label{eq:app03}
K\cdot{}_2F_2(1,1;2,2;-K)&=-E_i(-K)+\log K +\gamma_{e}\\&=\Gamma(0,K)+\log K +\gamma_{e},
\end{align}
where $\Gamma(a,x)$ denotes the upper incomplete gamma function, $\log$ is the natural logarithm and $\gamma_e$ is the Euler-Mascheroni constant. Finally, using the chain rule in (\ref{eq:app01}) and after some algebra, we have
\begin{equation}
\label{eq:app04}
\frac{\partial}{\partial n}\mathbb{E}(\gamma^{n})|_{n=0}=\Gamma(0,K)+\log K +\log \left(\frac{\bar\gamma}{K+1}\right).
\end{equation}
Expression (\ref{eq:app04}) is new in the literature to the best of our knowledge. Setting $K=0$, the scenario reduces to the Rayleigh case; using $\Gamma(0,K)+\log(K/(1+K))\rightarrow-\gamma_e$ as $K\rightarrow0$, we have
\begin{equation}
\label{eq:app04b}
\frac{\partial}{\partial n}\mathbb{E}(\gamma^{n})|_{n=0}=\log \bar\gamma -\gamma_e,
\end{equation}
which is coincident with the expression given in \cite[eq. 22]{Yilmaz2012b}. Similarly, if we let $K\rightarrow\infty$ in (\ref{eq:app04}), we have
\begin{equation}
\label{eq:app04c}
\frac{\partial}{\partial n}\mathbb{E}(\gamma^{n})|_{n=0}=\log \bar\gamma,
\end{equation}
i.e., the asymptotic capacity of the AWGN channel.
\section{Asymptotic capacity of GTR-U fading}
\label{app2}
Leveraging the result calculated in the previous appendix, we use lemma \ref{lemma1} to derive the asymptotic capacity of GTR fading in the high-SNR regime. Using the relationship
\begin{equation}
\label{eq:app05}
\frac{\partial}{\partial n}\mathbb{E}(\gamma^{n})|_{n=0}^{\text{GTR-U}}=\tfrac{1}{2\pi}\int_{0}^{2\pi}\frac{\partial}{\partial n}\mathbb{E}(\gamma^{n})|_{n=0}^{\text{Rice}}|_{K=K(1+\Delta\cos\theta)}d\theta,
\end{equation}
and noticing (\ref{eq:gamma}), we have
\begin{align}
\label{eq:app05b}
&\frac{\partial}{\partial n}\mathbb{E}(\gamma^{n})|_{n=0}^{GTR}=  \frac{1}{2\pi}\int_{0}^{2\pi}\log(1+\Delta\cos\theta)d\theta \nonumber\\&+\log \left(\frac{K \bar\gamma}{K+1}\right)+\frac{1}{2\pi}\int_{0}^{2\pi}\Gamma\left(0,K(1+\Delta\cos\theta)\right)d\theta.
\end{align}
The first integral can be expressed in closed-form as
\begin{align}
\label{eq:app06}
\frac{1}{2\pi}\int_{0}^{2\pi}\log(1+\Delta\cos\theta)d\theta= \log\left(\frac{1+\sqrt{1-\Delta^2}}{2}\right).
\end{align}
This term (\ref{eq:app06}) vanishes for $\Delta=0$, corresponding to the Rician case, whereas it takes the value $-\log 2$ for $\Delta=1$. Hence, its contribution to the asymptotic capacity is always negative for $\Delta\neq 0$, which is coherent with the observation that the parameter $\Delta$ is related with the severity of fading.

In order to solve the second integral, we first use the integral form for the incomplete gamma function that results from \cite[5.1.4]{Abramowitz1965}
\begin{align}
\label{eq:app07}
\Gamma(0,x)=\int_1^{\infty}\frac{e^{-xt}}{t}dt.
\end{align}
Hence, the integral of interest becomes
\begin{align}
\mathcal{J}(K,\Delta)=\frac{1}{2\pi}\int_{0}^{2\pi}\int_1^{\infty}\frac{e^{-tK(1+\Delta\cos\theta)}}{t}dtd\theta.
\end{align}
Changing the order of integration, we have
\begin{align}
\mathcal{J}(K,\Delta)=\int_1^{\infty} \frac{e^{-tK}}{t}\left\{\frac{1}{2\pi}\int_{0}^{2\pi}e^{-tK\Delta\cos\theta}d\theta\right\} dt,
\end{align}
where the inner integral can be identified as the modified Bessel function. Therefore, we have
\begin{align}
\label{eq:intapp}
\mathcal{J}(K,\Delta)=\int_1^{\infty} \frac{e^{-tK}}{t}I_0(tK\Delta) dt.
\end{align}
This integral can be efficiently computed numerically, and reduces to $\Gamma(0,x)$ for $\Delta=0$. However, since $t\geq 1$, we can obtain a very accurate approximation for (\ref{eq:intapp}) when $K\Delta>>1$ using the first term of the Hankel expansion of $I_0(z)=\tfrac{e^z}{\sqrt{2\pi z}}+\mathcal{O}(z^{-3/2})$ \cite[eq. 10.40.1]{Olver2010} to obtain
\begin{align}
\label{eq:intapp2}
\mathcal{J}(K,\Delta)&\approx\int_1^{\infty} \frac{e^{-tK}}{t}\frac{e^{tK\Delta}}{\sqrt{2\pi K t \Delta}} dt\\&=\sqrt{\frac{2}{\pi}}\left\{\frac{e^{-K(1-\Delta)}}{\sqrt{K\Delta}}-\sqrt{\left(\tfrac{1}{\Delta}-1\right)}\text{erfc}\left(K(1-\Delta)\right)\right\}\nonumber,
\end{align}
where $\text{erfc}(\cdot)$ is the complementary error function. For the particular case of $\Delta=1$, we have a very simple expression $\mathcal{J}(K,\Delta)\approx\sqrt{\frac{2}{\pi K}}$.
%

\bibliographystyle{ieeetr}
\bibliography{TWDP}

\begin{IEEEbiographynophoto}{Milind Rao} (S'14) received the B. Tech degree in Electrical Engineering from IIT Madras, India in 2013 where he graduated with the Siemens award for academic proficiency. He is currently pursuing the M.S. and Ph.D. degrees in Electrical Engineering at Stanford University under the supervision of Prof. Andrea Goldsmith. He is a recipient of the Stanford School of Engineering Graduate, DAAD and KVPY fellowships. His research interests are in stochastic control, optimization, and wireless communications.
\end{IEEEbiographynophoto}

\begin{IEEEbiographynophoto}{F. Javier Lopez-Martinez} (S'05, M'10) received the M.Sc. and Ph.D. degrees in Telecommunication Engineering in 2005 and 2010, respectively, from the University of Malaga (Spain). He joined the Communication Engineering Department at the University of Malaga in 2005, as an associate researcher. In 2010 he stays for 3 months as a visitor researcher at University College London. He is the recipient of a Marie Curie fellowship from the UE under the “U-mobility” program at University of Malaga. Within this project, between August 2012-2014 he held a postdoc position in the Wireless Systems Lab (WSL) at Stanford University, under the supervision of Prof. Andrea J. Goldsmith. He's now a Postdoctoral researcher at the Communication Engineering Department, Universidad de Malaga.
He has received several research awards, including the best paper award in the Communication Theory symposium at IEEE Globecom 2013. His research interests span a diverse set of topics in the wide areas of Communication Theory and Wireless Communications: stochastic processes, random matrix theory, statistical characterization of fading channels, physical layer security, massive MIMO and mmWave for 5G.
\end{IEEEbiographynophoto}

\begin{IEEEbiographynophoto}{Mohamed-Slim Alouini} (S'94, M'98, SM'03, F’09) was born in Tunis, Tunisia. He received the Ph.D. degree in Electrical Engineering
from the California Institute of Technology (Caltech), Pasadena, CA, USA, in 1998. He served as a faculty member in the University of Minnesota, Minneapolis, MN, USA, then in the Texas A$\&$M University at Qatar, Education City, Doha, Qatar before joining King Abdullah University of Science and Technology (KAUST), Thuwal, Makkah Province, Saudi Arabia as a Professor of Electrical Engineering in 2009. His current research interests include the modeling, design, and
performance analysis of wireless communication systems.
\end{IEEEbiographynophoto}

\begin{IEEEbiographynophoto}{Andrea Goldsmith} (S'90-M'93-SM'99-F'05) is the Stephen Harris professor in the School of Engineering and a professor of Electrical Engineering at Stanford University. She was previously on the faculty of Electrical Engineering at Caltech. Dr. Goldsmith co-founded and served as CTO for two wireless companies: Accelera, Inc., which develops software-defined wireless network technology for cloud-based management of WiFi  access points, and  Quantenna Communications, Inc., which develops high-performance WiFi chipsets. She has previously held industry positions at Maxim Technologies, Memorylink Corporation, and AT\&T Bell Laboratories. She is a Fellow of the IEEE and of Stanford, and has received several awards for her work, including the IEEE ComSoc Armstrong Technical Achievement Award,  the National Academy of Engineering Gilbreth Lecture Award, the IEEE ComSoc and Information Theory Society joint paper award, the IEEE ComSoc Best Tutorial Paper Award, the Alfred P. Sloan Fellowship, and the Silicon Valley/San Jose Business Journal’s Women of Influence Award. She is author of the book ``Wireless Communications'' and co-author of the books ``MIMO Wireless Communications'' and “Principles of Cognitive Radio,” all published by Cambridge University Press, as well as inventor on 25 patents. She received the B.S., M.S. and Ph.D. degrees in Electrical Engineering from U.C. Berkeley.

Dr. Goldsmith has served as editor for the IEEE Transactions on Information Theory, the Journal on Foundations and Trends in Communications and Information Theory and in Networks, the IEEE Transactions on Communications, and the IEEE Wireless Communications Magazine as well as on the Steering Committee for the IEEE Transactions on Wireless Communications. She participates actively in committees and conference organization for the IEEE Information Theory and Communications Societies and has served on the Board of Governors for both societies. She has also been a Distinguished Lecturer for both societies, served as President of the IEEE Information Theory Society in 2009, founded and chaired the Student Committee of the IEEE Information Theory Society, and chaired the Emerging Technology Committee of the IEEE Communications Society. At Stanford she received the inaugural University Postdoc Mentoring Award, served as Chair of Stanford’s Faculty Senate in 2009, and currently serves on its Faculty Senate, Budget Group, and Task Force on Women and Leadership. 
\end{IEEEbiographynophoto}

\end{document}